# Perovskite quantum dot topological laser

Jingyi Tian[1,2 †,*], Qi Ying Tan[1,3 †], Yutao Wang[1,3], Yihao Yang[1,2], Guanghui Yuan[1,2], Giorgio Adamo[1,2], Cesare Soci[1,2,*]

[1] *Centre for Disruptive Photonic Technologies, TPI, Nanyang Technological University, 21 Nanyang Link, Singapore 637371*

[2] *Division of Physics and Applied Physics, School of Physical and Mathematical Sciences, Nanyang Technological University, Singapore 637371*

[3] *Energy Research Institute @NTU (ERI@N), Interdisciplinary Graduate School, Nanyang Technological University, 50 Nanyang Drive, 637553, Singapore*

Correspondence: jingyi.tian@ntu.edu.sg, csoci@ntu.edu.sg

†These authors contributed equally to this work.

**Abstract**

Various topological laser concepts have recently enabled the demonstration of robust light-emitting devices that are immune to structural deformations and tolerant to fabrication imperfections. Current realizations of photonic cavities with topological boundaries are often limited by outcoupling issues or poor directionality and require complex design and fabrication that hinder operation at small wavelengths. Here we propose a topological cavity design based on interface states between two one-dimensional photonic crystals with distinct Zak phases and demonstrate a lithography-free, single-mode perovskite laser emitting in the green. Few monolayers of solution processed all-inorganic cesium lead halide perovskite quantum dot are used as ultrathin gain medium. The topological laser has planar design with large output aperture, akin to vertical-cavity surface-emitting lasers (VCSELs) and is robust against variations of the thickness of the gain medium, from deeply subwavelength to thick quantum dot films. This experimental observation also unveils the topological nature of VCSELs, that is usually overlooked in the description of conventional Fabry-Perot cavity lasers. The design simplicity and topological characteristics make this perovskite quantum dot laser architecture suitable for low-cost and fabrication tolerant vertical emitting lasers operating across the visible spectral region.

**Introduction**

The concept of topological photonics [1-3], characterized by global topological invariants of the photon wavefunction in the optical dispersion bands, has been used in recent years for the design of optical cavities that are robust against fabrication imperfections, leading to the demonstration of several topological micro/nanolasers operating in the infrared [4-12]. The cavity design of these lasers relies on the excitation of boundary states (e.g., edge or corner states) at the interface of photonic structures in different topological phases. The difference of topological invariants across the interface defines the boundary states while the so-called bulk-edge correspondence underpins their topological protection [1,13-16]. Typical edge/corner emitting topological lasers based on 1D chains [6] or 2D arrays [9] of semiconductors resonators bear poor directionality, are difficult to outcouple and are limited to infrared operation by their design complexity. Recently, topological lasers based on topological bulk bands [12] and topological vertical-cavity laser arrays [10] addressed the directionality and outcoupling issues, but did not improve on design complexity and access to short emission wavelengths. Here we demonstrate a lithography-free, single-mode topological laser with interface states confined between two 1D photonic crystals with distinct Zak phases (i.e., the geometric phase picked up by a photon moving across the one-dimensional Brillouin zone) [13,17]. Akin to vertical-cavity surface-emitting lasers (VCSELs) [18], the proposed planar design allows achieving large output apertures while retaining robustness against variations of the gain medium thickness. Solution-processed films of green-emitting halide perovskite quantum dots are used as gain media due to their large optical absorption cross section, large exciton binding energy, high photoluminescence quantum yield and low Anger recombination loss [18,19]. With gain media thickness varying from subwavelength to approximately half of the working wavelength, we probe experimentally the transition from an interface-state topological laser to a conventional Fabry-Perot cavity laser, unveiling the predicted topological nature of VCSELs [20].

**Results**

*Design principles*

The existence of localized optical states at the interface between two 1D binary photonic crystals (PCs) [21,22] is determined by their surface impedance mismatch. This is related to the geometric phases (i.e., the Zak phases) of the bulk optical bands lying below the PC bandgap on either side of the interface. The excitation of optical states confined near the interface is possible when the two PCs have surface impedances of opposite sign. In such systems with inversion symmetry, the Zak phase is a quantized topological invariant with value of 0 or $\pi$, corresponding to opposite parity of the optical modes with respect to the inversion center. For instance, the planar topological cavity depicted in Fig. 1a comprises of the interface between two semi-infinite 1D binary photonic crystals (PC1 and PC2) with inversion symmetry along the *z*-direction. The PCs consist of alternating low index (LI, with thickness of $d_1$) and high index (HI, with thickness of $d_2$) layers, at the middle of which lay the inversion centers of PC1 and PC2, respectively.

To minimize the cavity thickness, the topological cavity in our demonstration is designed to work within the first optical bandgap of a $TiO_2/SiO_2$ multilayer, with $n_{HI}$=2.3, $n_{LI}$=1.5, $d_1$=62 nm, $d_2$=70 nm. Although PC1 and PC2 share the same bulk optical bands, they carry distinct Zak phases for the lowest band due to the different placement of LI and HI about their inversion centers (Figs. 1b and 1c). Analytically, this is seen in the definition of the Zak phase $\theta_{Zak}$ carried by the lowest band [13]:

$$\exp(i\theta_{Zak}) = sign\left[1 - \frac{\mu_c \varepsilon_c}{\mu_s \varepsilon_s}\right] = sign(\eta), \tag{1}$$

Where $\eta$ is the surface impedance and $\varepsilon_c$, $\mu_c$ indicate the relative permittivity and permeability of the layer containing the inversion center and $\varepsilon_s$ and $\mu_s$ those of the side layers. From Eq. 1, PC1 and PC2 have Zak phase of $\pi$ and 0 and *odd* and *even* electric field distribution across their inversion centers (insets of Fig. 1b), respectively. Thus, surface impedances of PC1 and PC2 have opposite sign and an interface state exists within the first bandgap of the proposed 1D topological cavity.

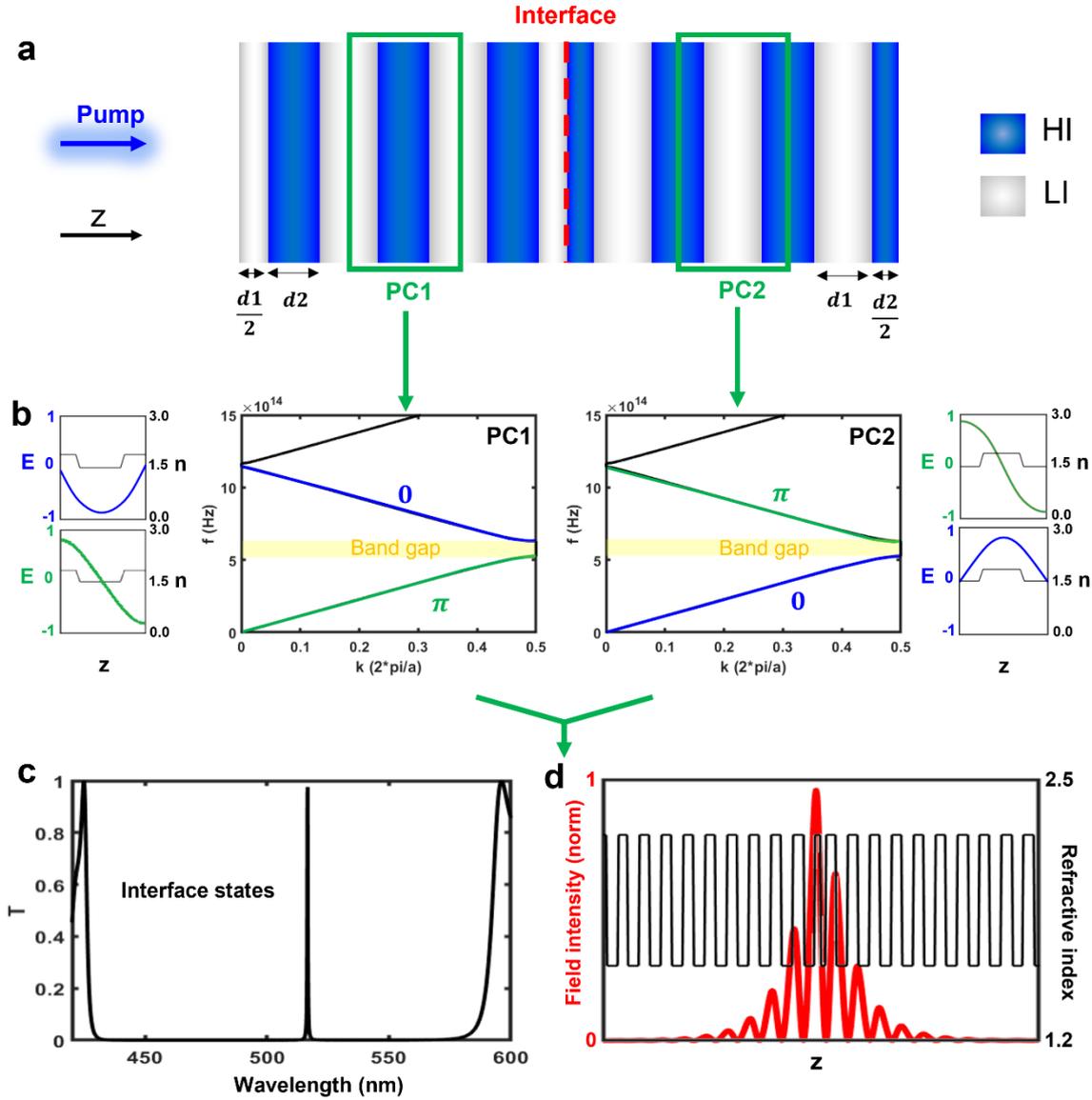

**Figure 1. Design principles of a 1D topological microcavity.** (a) Schematic of a one-dimensional (1D) topological microcavity formed by interfacing two photonic crystals (PC1 and PC2) composed of alternative high index (HI) and low index (LI) layers. The inversion center of PC1 is chosen to be within the LI layer, while the inversion center of PC2 is within the HI layer, as highlighted by the green boxes. (b) The calculated optical band structures of PC1 and PC2 are identical but have distinct Zak phases. The normalized electric field distributions along the z-axis within the respective unit cells for the first two optical bands are shown in the insets. (c) Calculated transmission spectra of the 1D topological cavity formed by 10 unit cells on each side of the interface, showing the appearance of a high-Q interface state within the optical bandgap. (d) Calculated spatial distribution of the electric field within the microcavity.

The 1D topological cavity, consisting of two half-cavities of 10 unit cells, is expected to show a high-quality interface state (Q > 5000) around 515 nm, at the center of the first optical bandgap, as predicted by the calculated transmission spectrum shown in Fig. 1c. The electric field is highly confined at the interface of the two PCs, with an asymmetric distribution peaking inside the first HI layer of PC1 (Fig. 1d).

We have experimentally verified the existence of the topological interface state by measuring the transmittance of a microcavity made of alternating layers of $TiO_2$ (n~2.3) and $SiO_2$ (n~1.5), shown in Fig. 2a. The measured transmission spectrum (Fig. 2b) is in good correspondence to the calculated one, with a high-quality factor interface state (Q~2000) appearing within the optical bandgap at λ=507.3 nm.

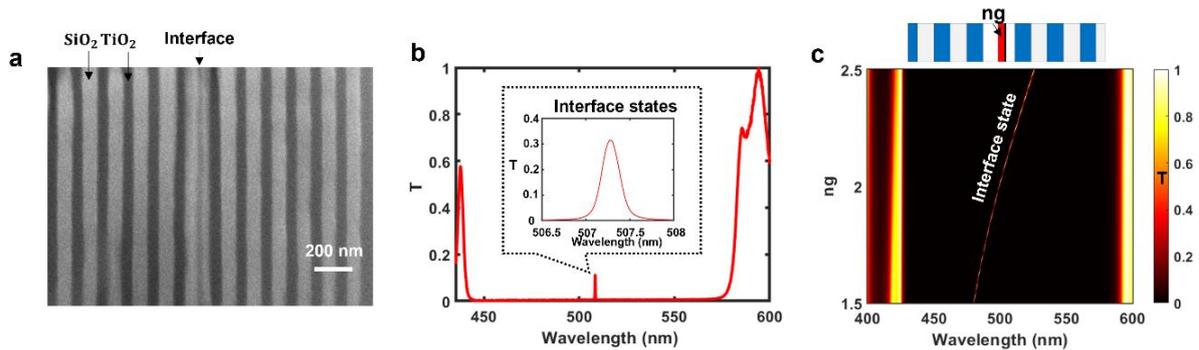

**Figure 2. Topological interface states in the microcavity.** (a) Cross-sectional scanning electron microscope (SEM) image of the 1D microcavity consisting of 20 alternating layers of $TiO_2$ and $SiO_2$ and on each side of the interface. (b) Measured transmission spectrum of the 1D topological cavity. The high-Q transmission peak (Q~2000) at 507.3 nm, arising from the interface state, is magnified in the inset. (c) Transmission spectra of the topological microcavity when a gain medium of refractive index varying from $n_g$=1.5 to $n_g$=2.5 replaces the first HI layer near the interface.

In view of employing this cavity design in a laser, we analyzed the robustness of the topological interface state against local perturbations of the lattice by substituting the first half HI layer with a hypothetical gain medium of refractive index $n_g$ varying from 1.5 to 2.5 (losses are neglected in first approximation). The corresponding transmission spectra are shown in the colormap of Fig. 2c. We find that, when the refractive index of the gain medium layer increases, the interface state continues to exist within the optical bandgap while undergoing a redshift from λ=480 nm to λ=520 nm. Remarkably, the spatial localization of the interface state is hardly influenced by the lattice perturbation (Fig. S1). Thus, spectral and spatial robustness of the topological state against variation of the refractive index at the interface make this cavity design ideal for a laser where the first half HI layer is substituted by a gain medium, provided that sufficient overlap between the optical gain and the topological mode is ensured.

*Lasing in the topological interface state*

To realize the topological laser, we used solution-processed all-inorganic cesium lead bromide (CsPbBr$_3$) quantum dot films to substitute the half HI interface layer of PC1. CsPbBr$_3$ quantum dots are highly efficient and cost-effective gain media [18,19]. We synthesized nanocrystals with nominal size of ~9 nm and spontaneous emission centered around λ=515 nm (Fig. 3a) and cast films with thickness *d*=40 nm and refractive index n~2 (Fig. S2) on the PC2 side, then mechanically bonded to the complementary PC1 side. The performance of the complete laser structure was characterized under λ=400 nm frequency-doubled fs-laser pump, with 100 fs pulse duration and 1 kHz repetition rate: the dependence of emission intensity on pump fluence is shown in Fig. 3b. The broad emission spectrum observed at low pump fluences is overtaken by a single, narrow lasing peak (λ=525.7 nm, FWHM = 0.4 nm) at higher pump fluences (Fig. 3b), indicating transition from the spontaneous to the stimulated emission regime. The S-shaped light-light curve, accompanied by more than one order of magnitude narrowing of the linewidth, is consistent with laser emission, with corresponding lasing threshold of 8.95 µJ/cm$^2$ (Fig. 3c). We experimentally confirmed that the topological laser maintains single mode operation (from λ=531.5 nm to λ=519) while varying the gain medium thickness from $d_g$=43 nm to $d_g$=37 nm (Fig. 3d). We note that the perovskite quantum dot topological laser has comparable lasing performance to a previously reported multimode VCSEL, which employed a hundred times thicker CsPbBr$_3$ nanocrystal film ($d_g$=4 μm) as the gain medium [19].

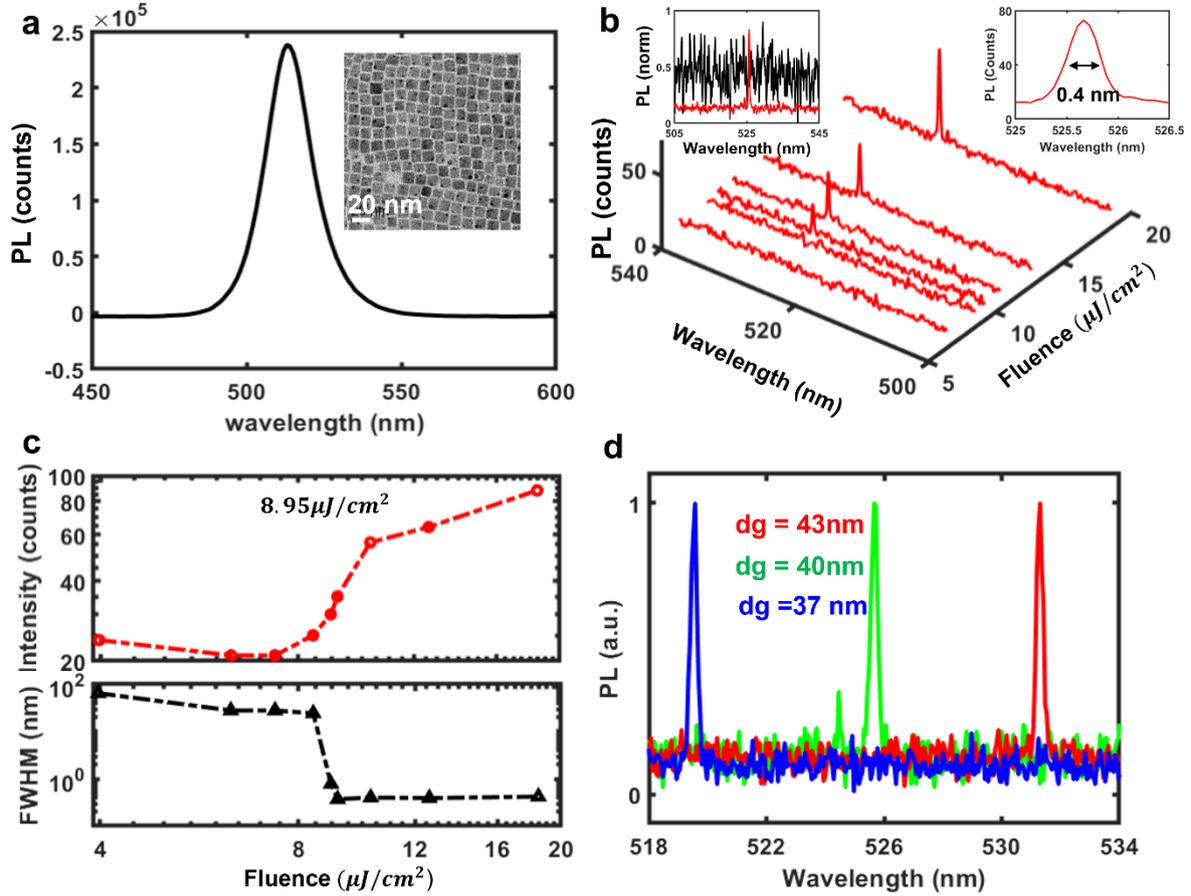

**Figure 3. Laser performance of the 1D topological microlaser at room temperature.** (a) Room temperature photoluminescence spectrum of the CsPbBr$_3$ quantum dots film. The inset shows a transmission electron microscopy image of the cuboid CsPbBr$_3$ quantum dots, with an average size of 9 nm. (b) Room-temperature emission spectra of the topological microlaser as function of pump fluence, showing laser peaks forming above threshold. The inset on the left shows the rescaled emission spectra below (black) and above (red) lasing threshold; the inset on the right shows the narrow linewidth of the laser peak obtained at pump fluence of 18.47 μJ/cm$^2$. (c) Light-light curve of the microlaser and linewidth of the emission spectra as a function of the pump fluence at room-temperature. (d) Dependence of the single mode laser emission wavelength on the perovskite quantum dot film thickness, $d_g$.

*From topological interface-state cavity lasers to VCSELs*

It has been theoretically suggested that the cavity modes of VCSELs [18,19,23] are mathematically equivalent to topological surface or edge states [20]. The thickness of the gain medium (including a spacer layer [24]) in such a Fabry-Perot cavity laser, however, cannot be indefinitely thin as it is bound to be an integer multiple $N$ of half the emission wavelength ($d = \frac{\lambda_g}{2} N$, $N = 1,2,3…$) to fulfill the condition of constructive interference of the resonance in the cavity. Thus, the practical realization of an actual topological interface state in physically thick Fabry-Perot cavities, remains puzzling. Our multilayer structure allows realizing the transition from a deeply subwavelength interface-state topological cavity to a rather thick Fabry-Perot cavity and experimentally verify the theoretical assumption.

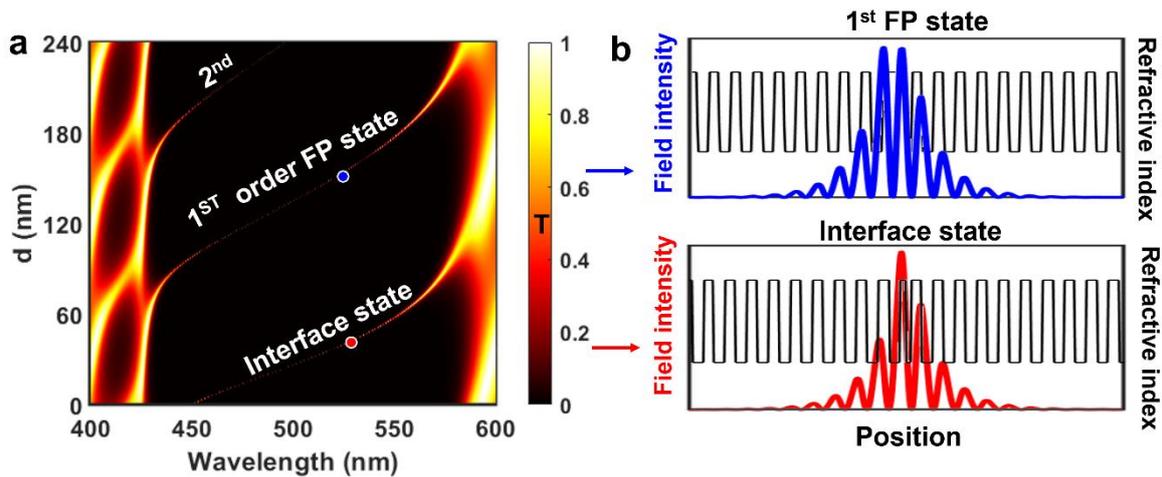

**Figure 4. Topological nature of conventional vertical cavity surface emitting lasers based on Fabry-Perot resonances.** (a) Calculated transmission spectra of the 1D topological cavity as function of the thickness of the first HI layer next to the interface (*d*). (b) Electric field distribution of the interface state and 1st order Fabry-Perot (λ/2-cavity) resonance inside the microcavity.

When plotting the calculated transmission spectra of the topological cavity as function of the first HI layer thickness, *d* (Fig. 4a), we find that the interface state survives within the optical bandgap up to *d*~60 nm, barring a wavelength shift. For larger values of *d*, the topological cavity becomes effectively equivalent to a 1st order Fabry-Perot cavity, where the cavity state also persists for a wide range of thicknesses, up to $d$~180 nm. The electric field distributions for the topological interface state (red circle in Fig. 4a) and the 1$^{st}$ order Fabry-Perot state (blue circle in Fig. 4a) are shown by the corresponding red and blue curves in Fig. 4b, along with the refractive index profiles of the cavity. Both show enhancement and confinement of the electric field around the layer where the gain medium is placed.

To compare the lasing action of the 1$^{st}$ order Fabry-Perot state experimentally, we embedded a 150 nm thick CsPbBr$_3$ quantum dot film and pumped the laser cavity under the same conditions used for the topological interface state laser, at room temperature. The corresponding laser performance is illustrated in Fig. S3, yielding comparable lasing threshold of 8 µJ/cm$^2$ and somehow broader emission linewidth of 0.8 nm, which can be attributed to the larger roughness of the thick quantum dot film. The observation of a seamless transition between the deeply subwavelength interface-state topological laser and the VCSEL provides a clear experimental confirmation that Fabry-Perot cavity lasers and topological lasers fall under the same category, as theoretically predicted [20].

**Discussion**

The concept of topology, which in recent years has branched from solid-state materials to photonics, has provided a powerful degree of freedom for the manipulation of light and has been successfully used to design robust optical cavities underpinning topological micro- and nanolasers operating in the infrared.

In this work, we have proposed a topological cavity design supporting interface states between two 1D photonic crystals with distinct Zak phases and experimentally realized a lithography-free CsPbBr$_3$ quantum dot single-mode laser emitting in the visible. The 1D topological cavity, where the thickness of the gain medium can be arbitrarily small, is particularly suitable for the realization of quantum dots lasers, as it allows overcoming issues of poor uniformity, clustering and luminescence quenching typical of quantum dot films. Furthermore, its planar configuration, akin to vertical-cavity surface-emitting lasers, allows achieving a large output aperture while showing robustness against perturbations in the multilayer architecture: without changing the structure, the thicknesses of the perovskite gain medium could be varied from deeply subwavelength to λ/2, equivalent to a Fabry-Perot cavity, unveiling the topological nature of VCSELs. Since electrically driven perovskite light-emitting devices with compatible stacked architecture are well established, this topological cavity design may provide a foundation for the highly sought electrically driven perovskite laser.

**Materials and methods**

*1. Synthesis of CsPbBr$_3$ quantum dots*

Cesium carbonate (Cs$_2$CO$_3$, 99.9% trace metals basis), lead (II) bromide (PbBr$_2$, 98%), oleylamine (OLA technical grade, 70%) and n-octane (anhydrous, ≥99%) were purchased from Sigma-Aldrich. Octadecene (ODE, technical grade 90%) and oleic acid (OA, technical grade 90%) were purchased Alfa Aesar. For the synthesis of Cs-oleate, 0.326 g of Cs$_2$CO$_3$, 18 mL of ODE and 1 mL of OA were loaded into a 100 mL three-neck flask. The mixture was heated under vacuum at 100°C for an hour and subsequently raised to 150°C under nitrogen flow. Upon complete dissolution of Cs$_2$CO$_3$, the solution was kept at 150°C to avoid solidification. For the synthesis of CsPbBr$_3$ quantum dots, 0.067g of PbBr$_2$, 5 mL of ODE, 0.5 mL of OA and 1 mL OLA were loaded into a separate 100 mL three-neck flask. The mixture was heated under vacuum at 100°C for an hour and subsequently raised to 150°C under nitrogen flow. After the complete dissolution of PbBr$_2$, the solution was further heated to 170°C. 0.6 mL of the as-prepared Cs-oleate was quickly injected and the solution was immediately placed into an ice – water bath and cooled to room temperature. To purify the resulting solution, ethyl acetate was subsequently added into the solution and the solution was centrifuged at 6000 rpm for 5 minutes. The precipitate was collected and dispersed in n-octane. The solution was again centrifuged at 6000 rpm for 5 minutes and the supernatant was in a nitrogen filled environment to prevent degradation of the solution.

## 2. *Structural and morphological characterization*

Transmission electron microscopy images were acquired on a JEM-1400 flash electron microscope operating at 100 kV. Samples were prepared by drop casting the solution onto a 400 mesh copper grids with carbon supporting films. Atomic force microscopy images were acquired on a Cypher ES microscope. Samples were prepared by spin coating the solution onto a glass microscope substrate.

## 3. *Device fabrication*

The topological DBR mirrors were cleaned sequentially with acetone, 2-propanol and deionized water under sonication for 5 minutes each. For the fabrication of the quantum dots films, the $CsPbBr_3$ quantum dots solution was spin-coated onto the cleaned DBR mirror at 3500 rpm for 60 seconds. Subsequently, the other DBR mirror was bonded to the as-coated mirror with epoxy applied at the edges and pressed with a heavy load to reduce the presence of the air gaps. The device was left to dry in a nitrogen filled environment for 24 hours.

## 4. *Device characterization*

Photoluminescence of the metasurface is measured by a frequency doubled Ti:Sapphire laser (400 nm, using a BBO crystal) from a regenerative amplifier (repetition rate 1 kHz, pulse width 100 fs, seeded by Mai Tai, Spectra Physics). The pumping laser is focused by a convex lens (with focus length of 3 cm) onto the top surface of the sample and the spot size on the sample is about 50 µm. Emitted light and corresponding fluorescence microscopy image are collected on the backside of the metasuface by a 5X objective lens coupled with a CCD coupled spectrometer (Acton IsoPlane SCT 320) and a camera, respectively. An attenuator and an energy meter are used to tune and measure the pumping density.

**Acknowledgements**

We would like to thank Elena Feltri for her help in optimizing film deposition and bonding. Research was supported by the A*STAR-AME programmatic fund on Nanoantenna Spatial Light Modulators for Next-Gen Display Technologies (Grant A18A7b0058), the Singapore Ministry of Education (MOE2016-T3-1-006), and the Quantum Engineering Programme of the Singapore National Research Foundation (NRF2021-QEP2-01-P01).

**Author contributions**

J.T., Y.Y., G.A. and C.S. conceived the idea. J.T. performed the numerical simulations and theoretical analysis, designed the experiments with G.A. and G.Y, and performed the laser measurements with Y.W. Q.Y.T. and J.T. synthesized the perovskite QD film and fabricated the microlaser samples. J.T., G.A. and C.S. analyzed the data and drafted the manuscript. All the authors contributed to finalizing the manuscript. C.S. supervised the work.

**Competing financial interests**

The authors declare no competing financial interests.

**Additional information**

**Supplementary information** is available in the online version of the paper.

**Data availability**

The authors declare that all data supporting the findings of this study are available within this article and its supplementary information and are openly available in NTU research data repository DR-NTU (Data) at https://doi.org/XXXXXX. Additional data related to this paper may be requested from the authors.

# Supporting Information

**Perovskite quantum dot topological laser**

Jingyi Tian[1,2 †,*], Qi Ying Tan[1,3 †], Yutao Wang[1,3], Yihao Yang[1,2], Guanghui Yuan[1,2], Giorgio Adamo[1,2], Cesare Soci[1,2,*]

[1] Centre for Disruptive Photonic Technologies, TPI, Nanyang Technological University, 21 Nanyang Link, Singapore 637371

[2] Division of Physics and Applied Physics, School of Physical and Mathematical Sciences, Nanyang Technological University, Singapore 637371

[3] Energy Research Institute @NTU (ERI@N), Interdisciplinary Graduate School, Nanyang Technological University, 50 Nanyang Drive, 637553, Singapore

Correspondence: jingyi.tian@ntu.edu.sg, csoci@ntu.edu.sg

†These authors contributed equally to this work.

## I. BAND STRUCTURE OF BINARY ONE-DIMENSIONAL PHOTONIC CRYSTALS

The band structures of a single binary PC can be calculated as follows,

$$\cos(k(d_1 + d_2)) = \cos k_1 d_1 \cos k_2 d_2 - \frac{1}{2}\left(\frac{z_1}{z_2} + \frac{z_1}{z_2}\right)\sin k_1 d_1 \sin k_2 d_2, \quad (1)$$

where $k_i = \omega n_i/c$, $n_i = \sqrt{\mu_i \varepsilon_i}$, $z_i = \sqrt{\mu_i/\varepsilon_i}$, (i = 1, 2 ), $k$ is the Bloch wave vector, $c$ denotes the speed of light in vacuum, and $\varepsilon_i$ and $\mu_i$ are the relative permittivity and permeability in the corresponding media, respectively.

## II. ROBUSTNESS OF INTERFACE STATE IN THE 1D TOPOLOGICAL MICROCAVITY

The robustness of the topological interface state can be tested inducing local perturbations in the lattice, particularly to the layers closest to the interface, where the electric field of the topological state is strongly localized. The transmission spectra in the colormap of Fig. S1a demonstrate the robustness of the interface state upon change of the refractive index $n_g$ of the first HI layer from 1.5 to 2.5. With the increase of $n_g$, the interface state continues to exist within the optical band gap while undergoing a redshift from $\lambda=480$ nm to $\lambda=520$ nm. Remarkably, the spatial localization of the interface state is hardly influenced as shown in Fig. S1b.

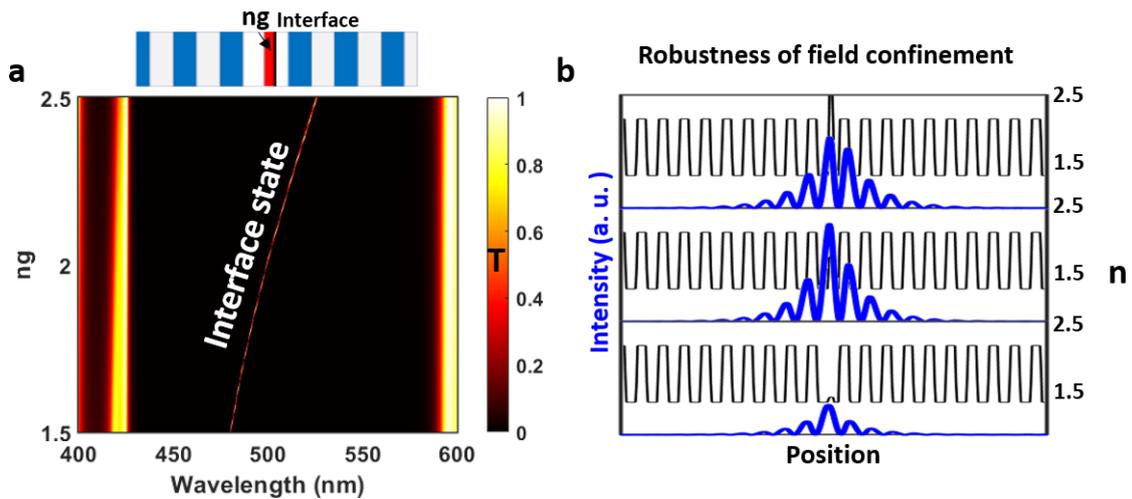

**Fig. S1 (a)** Calculated transmission map when the refractive index $n_g$ of the first HI layer near the interface is changed from 1.5 to 2.5. **(b)** Electric field distribution along the z-axis within the microcavity at the topological interface state when $n_g = 1.5$, $n_g = 2.0$ and $n_g = 2.5$, respectively.

# III. STRUCTURAL AND MORPHOLOGICAL PROPERTIES OF CsPbBr₃ QUANTUM DOTS

The morphological features of the CsPbBr$_3$ quantum dot film are shown in Fig. S2. *The cuboid CsPbBr$_3$ quantum dots have an average size of 9 nm (Fig. S2a). The as-synthesized CsPbBr$_3$ quantum dot film shows a thickness of 40 nm with a root mean square roughness of 8.6 nm (Fig. S2b and Fig. S2c).*

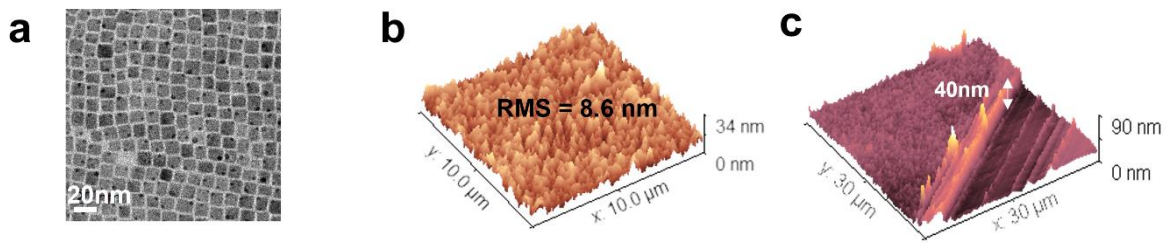

**Fig. S2** (a) Transmission electron microscopy image of the CsPbBr$_3$ quantum dots. (b-c) Atomic force microscopy image of the fabricated CsPbBr$_3$ quantum dot film showing (b) a root mean square roughness of 8.6 nm and (c) a thickness of 40 nm.

## IV. PERFORMANCE OF VERTICAL-CAVITY SURFACE-EMITTING LASER

*Lasing at room temperature from a conventional vertical-cavity surface-emitting laser (VCSEL) with thickness of gain medium of 150 nm is characterized under frequency-doubled fs-laser pump ($\lambda$ = 400 nm), with 100 fs pulse duration and 1 kHz repetition rate. The pump-fluence dependence shown in Fig. S3. The broad emission spectrum observed at low pump fluences is overtaken by a single narrow lasing peak ($\lambda$ = 523.7 nm, FWHM = 0.8 nm) at higher fluences (Fig. s3a), with a lasing threshold of 8 µJ/cm$^2$ derived by the onset of the light-light curve (Fig. s3b). The VCSEL yields a lasing threshold similar to that of the microlaser.*

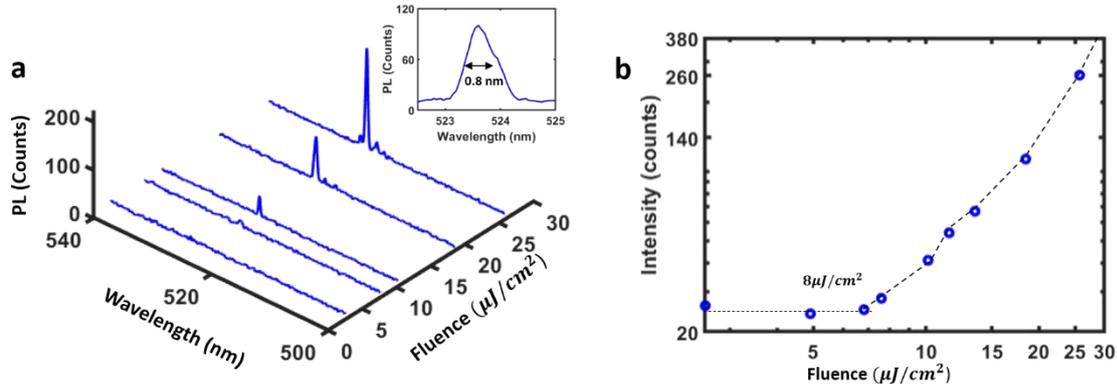

**FIG. S3** (a) Room-temperature emission spectra of *a conventional vertical-cavity surface-emitting laser (VCSEL)* as function of pump fluence. The inset shows the rescaled lasing emission spectra, indicating a linewidth of 0.8 nm. (b) Light-light curve of the VCSEL at room-temperature.